\documentclass[10pt,twocolumn]{article}

\usepackage[utf8]{inputenc}
\usepackage[T1]{fontenc}
\usepackage{lmodern}
\usepackage{geometry}
\usepackage{graphicx}
\usepackage{amsmath,amssymb}
\usepackage{booktabs}
\usepackage{microtype}
\usepackage{url}
\usepackage{hyperref}
\usepackage{xcolor}
\usepackage{caption}
\usepackage{subcaption}

\usepackage[table]{xcolor}
\definecolor{bestgreen}{HTML}{C3E6CB}    
\definecolor{secondgreen}{HTML}{E2F0D9} 

\usepackage{booktabs}   
\usepackage{multicol}   

\title{\bfseries audio2chart: End to End Audio Transcription into playable Guitar Hero charts}
\author{
  Riccardo Tripodi
}
\date{}  

\begin{document}

\twocolumn[
\maketitle
\begin{abstract}
This work introduces \textit{audio2chart}, a framework for the automatic generation of \textit{Guitar Hero style} charts directly from raw audio. 
The task is formalized as a sequence prediction problem, where models are trained to generate discrete chart tokens aligned with the audio on discrete time steps. 
An unconditional baseline demonstrates strong predictive performance, while the addition of audio conditioning yields consistent improvements across accuracy based metrics. 
This work demonstrates that incorporating audio conditioning is both feasible and effective for improving note prediction in automatic chart generation. 
The complete codebase for training and inference is publicly available at \href{https://github.com/3podi/audio2chart/tree/main\#}{\texttt{github.com/3podi/audio2chart}} supporting reproducible research on neural chart generation. A family of pretrained models is released on Hugging Face at \href{https://huggingface.co/3podi}{\texttt{huggingface.co/3podi}}.
\end{abstract}
\vspace{1em}
]

\vspace{1em}

\section{Introduction}
Rock ‘n’ roll never dies. Despite the decline of the mainstream video game series that popularized rock-based rhythm games, the online community continues to keep the spirit alive through free-to-play replicas. A key factor in sustaining these games is the ability for any user to add custom charts for their favorite songs. However, while the community works hard to create as many playable charts as possible, the process of charting is time-consuming and requires specialized skills that not everyone has, or is willing to learn, limiting chart creation to a small group of experienced users. The core challenge involves analyzing audio signals to detect onset times, beat positions, and musical structure while simultaneously considering gameplay factors such as difficulty progression, hand movement patterns, and player engagement. Notes must not only align with the audio but also create enjoyable and physically feasible gameplay sequences.

\vspace{1em}
From a deep learning perspective, the charting task can be framed as a sequence-to-sequence problem, where the input is the audio track and the output is a sequence of notes with their corresponding temporal positions. In this work, I present an approach to automating the creation of \textit{Guitar Hero style} charts by training neural network models capable of learning the audio to chart mapping directly from data.

\section{Related Work}
Neural network architectures for rhythm game chart generation have evolved in close connection with advances in music information retrieval, automatic music transcription, and sequence modeling. While research on generating playable charts has focused on different games such as \textit{Beatmania}, \textit{Dance Dance Revolution}, and \textit{Osu!} these efforts share a common foundation in audio feature analysis and sequence modeling.

\vspace{1em}

Early studies employed relatively simple feed-forward networks with multiple fully connected layers to predict note placements. For example, in \textit{Beatmania} chart generation, a multi-layer network with ReLU activations was trained to identify which sounds should be mapped to player actions \cite{lin2019generationmanialearningsemanticallychoreograph}. However, the inherently temporal and hierarchical nature of musical structure soon led researchers toward more advanced sequence modeling approaches.

Hybrid CNN-RNN architectures became the dominant paradigm for speech processing \cite{amodei2015deepspeech2endtoend}, combining the spatial feature extraction of convolutional networks with the temporal modeling capacity of recurrent networks. The influential work \cite{donahue2017dancedanceconvolution} introduced an LSTM-CNN encoder model adapted from speech recognition, demonstrating that such architectures could effectively process spectrogram representations to predict step placements for \textit{Dance Dance Revolution}.

More recently, Transformer-based architectures \cite{vaswani2023attentionneed} have established themselves as a highly general and flexible framework for sequence-to-sequence problems. The encoder-decoder design, built on self-attention mechanisms, allows these models to learn complex mappings between input and output sequences without requiring task-specific inductive biases, making them applicable to a wide variety of domains including audio processing and speech recognition \cite{radford2022robustspeechrecognitionlargescale}. Their ability to capture long-range dependencies and hierarchical structure in sequential data has positioned Transformers as a unifying architecture for diverse seq2seq tasks. In the domain of rhythm game chart generation for \textit{Osu!}, Yi et al. employed an encoder-decoder Transformer that takes log-mel spectrograms as input and autoregressively generate sequences of chart tokens, leveraging beat-aligned conditioning and difficulty embeddings \cite{yi2023beatalignedspectrogramtosequencegenerationrhythmgame}.

To the best of my knowledge, there is currently no working solution for automatically charting \textit{Guitar Hero style} songs.

\section{Dataset}

\begin{figure*}[h]
\centering
\includegraphics[width=0.9\textwidth]{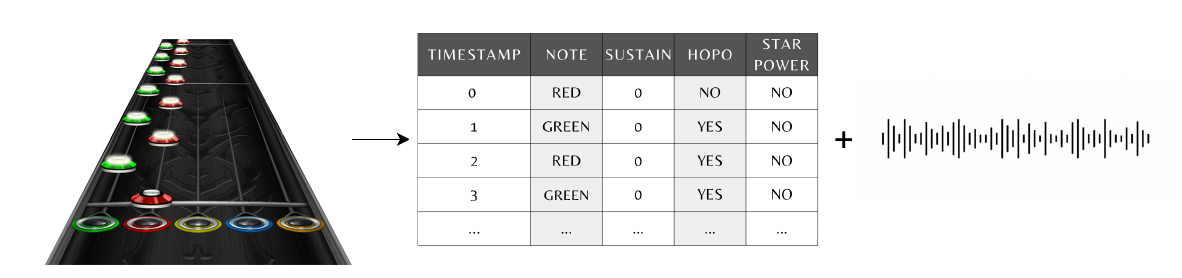}
\caption{Example of a dataset sample. Each gameplay song is represented by its audio track and a sequence of note events, each annotated with additional attributes such as note duration, HOPO, star power, and other in-game mechanics. For simplicity the timestamp here is represented as an integer number but in practice it can be any real number since a note can be placed at any time. Notes happening at the same time are represented as different rows with the same timestamp.}
\label{fig:dati}
\end{figure*}

The dataset comprises approximately 10{,}000 paired audio tracks and their corresponding \textit{Guitar Hero}--style charts. Each chart encodes a sequence of note events aligned with musical time, specifying both the onset and duration of each note within the song. In addition to timing information, each note includes categorical attributes describing gameplay mechanics such as hammer-ons and pull-offs (HOPOs), tap notes, and star power sections, as illustrated in Figure~\ref{fig:dati}.

Focusing on the most commonly played configurations, there are four supported instruments (Lead Guitar, Bass Guitar, Drums, and Keys) and four difficulty levels: Expert, Hard, Medium, and Easy. 




The dataset is strongly unbalanced, with a clear predominance of \textit{Expert} charts. This class not only appears most frequently but also contains the highest number of notes per song and the highest note density (Table~\ref{table:notes_per_song}). Such an imbalance makes it more challenging to train a single multiclass model across difficulty levels, as the model is naturally biased toward the characteristics of the Expert class.

\begin{table}[h]
\centering
\setlength{\tabcolsep}{4pt} 
\begin{tabular}{lccc}
\toprule
\textbf{Diff.} & \textbf{\%} & \textbf{Notes/Song} & \textbf{Notes/min} \\
\midrule
Expert & 70.4 & 1285 & 294.8 \\
Hard   & 10.4 & 823  & 206.1 \\
Medium & 9.9  & 579  & 145.5 \\
Easy   & 9.3  & 420  & 105.4 \\
\bottomrule
\end{tabular}
\caption{Distribution of difficulty levels in the dataset, along with the average number of notes per song and per minute for each level.}
\label{table:notes_per_song}
\end{table}



The distribution of songs duration is similar among the difficulty levels with an average duration of 3.9 minutes per song (excluding outliers because the community likes to chart full albums in a single audio track and chart file).



Check appendix \ref{appendix_dataset} for more information about the dataset.

\subsection{Tokenization}
In \textit{Guitar Hero}, gameplay simulates guitar performance through a simplified input interface consisting of five fret buttons (colored green, red, yellow, blue, and orange) and a strum bar that functions as the picking mechanism. Each note in the game corresponds to one or more of these fret buttons, which can be pressed individually or in combination to form chords.

In the raw chart files, notes are represented as a sequence of timed events, where chords correspond to multiple note events occurring at the same timestamp. To prepare these charts for neural network processing, it is necessary to tokenize the sequences in a way that preserves their temporal and harmonic structure. The most natural approach is to group together all note events that occur simultaneously into a single token.

In practice, the system allows for five single-button notes plus an open note (no fret pressed). Considering all possible combinations of these six states, including multi-button chords, there are 63 distinct note configurations in total.

Nevertheless, certain higher-order combinations occur far less frequently than others. For instance, although the game mechanics allow for combinations involving the open note alongside other fret buttons, such configurations appear exclusively in the \textit{Expert} difficulty level and constitute the extreme tail of the token frequency distribution, with occurrence rates ranging from approximately 0.001\% down to $10^{-6}\%$. So this cases must be carefully handled during training or completely omitted.

Overall, this tokenization method resemble a char-level tokenization. Higher order tokenization may be possible aggregating together the most frequent pairs of tokens in BPE style \cite{10.5555/177910.177914}. This would exclude the possibility to predict at the same time notes and their timestamp or additional attributes and would require a second step of learning for predicting the position in time of expanded sequences.

Check appendix \ref{appendix_tokenization} for the full notes-tokens mapping and tokens distributions.

\section{Method}

The task of \textit{Guitar Hero} chart generation can be formulated as an autoregressive sequence modeling problem. Given a sequence of discrete tokens representing notes or chords, $\mathbf{y} = (y_1, y_2, \dots, y_T)$, the model learns to predict the next token in the sequence conditioned on the previous context and on external information $\mathbf{c}$. This can be expressed as:
\begin{equation}
    P(\mathbf{y}) = \prod_{t=1}^{T} P(y_t \mid y_{<t}, \mathbf{c}),
\end{equation}
where $y_{<t} = (y_1, \dots, y_{t-1})$ denotes the previously generated tokens and $\mathbf{c}$ represents the conditioning variable, which may correspond to an audio embedding or to a learned difficulty-level embedding. The conditioning allows the model to adapt its predictions to both the musical content and the gameplay difficulty.

\vspace{0.5em}
A neural network parameterized by $\theta$ models this conditional distribution by mapping the past context and conditioning information to a set of output predictions:
\begin{equation}
    P_\theta(y_t \mid y_{<t}, \mathbf{c}) = g_\theta(f_\theta(y_{<t}, \mathbf{c})),
\end{equation}
where $f_\theta$ represents the network’s internal representation of the current state and $g_\theta$ denotes a task-specific output transformation that produces a valid probability distribution or regression output, depending on the type of prediction head.

\vspace{0.5em}
In practice, each time step of the sequence may require predicting more than a single discrete token. Each note event may also carry additional information, such as its duration or type (e.g., tap note, hammer-on/pull-off, or star power). To account for this, the model can be designed with multiple output heads, where each head predicts a different attribute of the current time step. The main head predicts the token identity, while auxiliary heads predict continuous or categorical properties associated with the same event.

\vspace{0.5em}
Each prediction head can have its own loss function, depending on the nature of the target variable. The token head is trained using a categorical cross-entropy loss:
\begin{equation}
    \mathcal{L}_{\text{token}} = - \sum_{t=1}^{T} \log P_\theta(y_t \mid y_{<t}, \mathbf{c}),
\end{equation}
The overall training objective combines the main categorical loss with the auxiliary heads objectives as:
\begin{equation}
    \mathcal{L} = \mathcal{L}_{\text{token}} + \sum_i \lambda_i \mathcal{L}_{\text{aux},i},
\end{equation}
where $\lambda_i$ are weighting coefficients controlling the relative importance of each auxiliary term.

\vspace{0.5em}
This multi-head design allows the model to jointly learn the discrete structure of note sequences and their continuous properties.

A fundamental aspect of chart generation lies in modeling both the \textit{sequence of notes} and their \textit{temporal placement}. While sequence modeling naturally follows an autoregressive formulation, where each token depends on the previous context, the precise positioning of notes in time remains an open challenge. Two main end-to-end strategies can be employed to jointly learn sequence and timing information. The first approach discretizes time into uniform intervals (e.g., every $N$ milliseconds), predicting a token at each time step along a fixed grid. This formulation allows time alignment to emerge directly from the autoregressive process but requires introducing special tokens for silent intervals, thus increasing sequence length. The second approach treats temporal information as an additional continuous attribute associated with each event, predicted by a dedicated time-placement head alongside the token identity. In this case, sequences are shorter and more compact, but the model must learn accurate timing through joint optimization of the token and time heads. Alternatively, the two tasks can be decoupled and handled by separate models trained sequentially, where one model predicts note sequences and another refines or aligns their temporal positions.

In any case during inference, a model can generate the sequence autoregressively, sampling one token at a time. The resulting sequence can then be decoded back into a playable chart representation mapping the tokens back to notes and chords.

In this work, we adopt the time-discretized tokenization strategy described in more detail in the following Section.

\subsection{Time Discretization}

\begin{figure}[htbp]
    \centering
\includegraphics[width=\columnwidth,height=0.4\textheight,keepaspectratio]{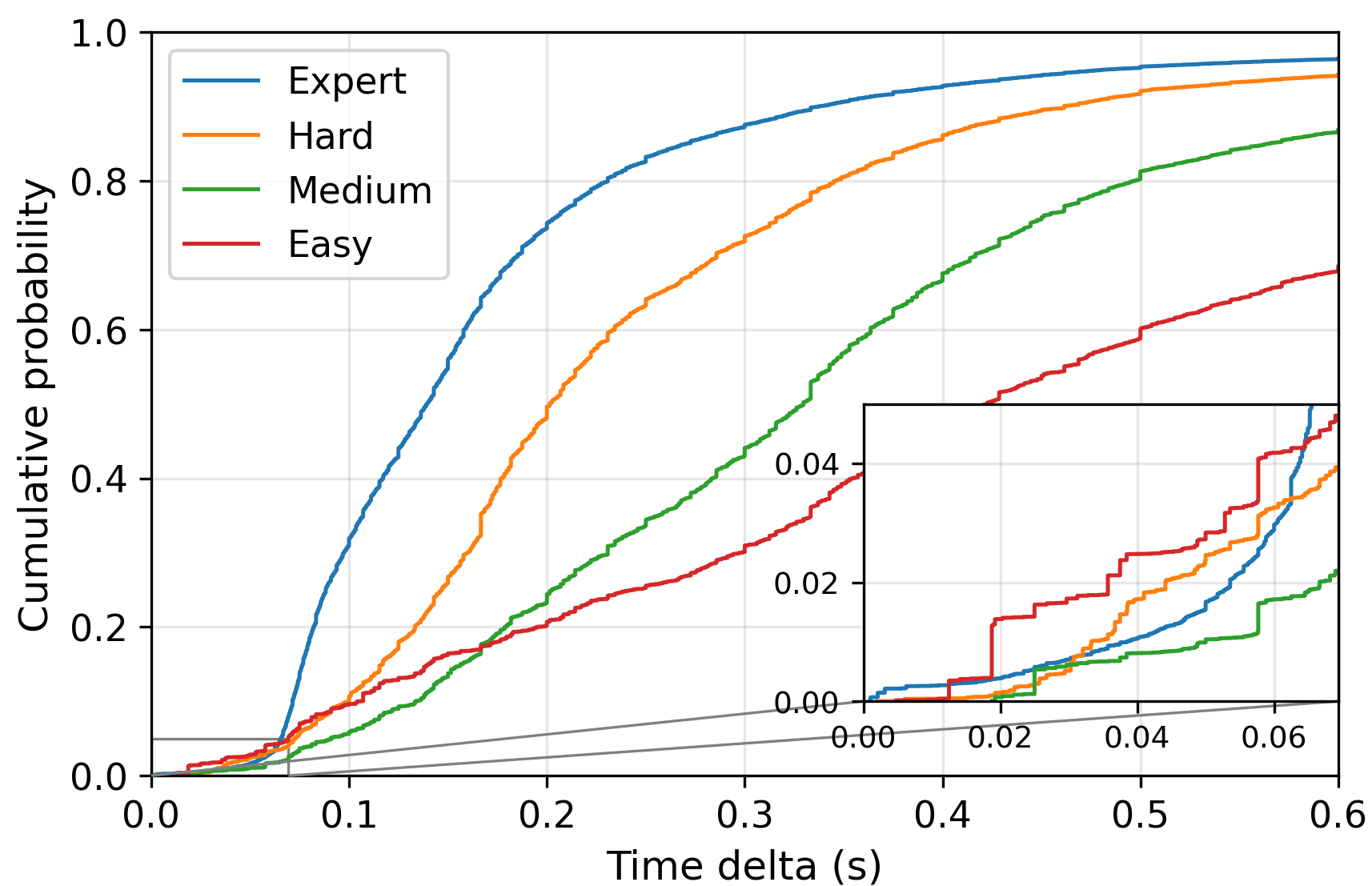}
    \caption{Cumulative distribution function of the time difference between consecutive notes for each difficulty level. Zoom on the first 0.5 seconds.}
    \label{fig:cdf}
\end{figure}

In the discrete time formulation, the encoded notes sequences are divided into uniform time steps, and a \textit{pad token} is inserted at time instants without any corresponding musical event. This setup yields a more challenging prediction problem: the finer the temporal resolution, the longer the sequences and the higher the proportion of pad tokens relative to meaningful musical tokens. Consequently, the model must implicitly learn to distinguish between \textit{silent} and \textit{non-silent} time instants.

A critical design choice in this approach is the time grid spacing, i.e., the temporal resolution used to discretize the events. A resolution that is too coarse may merge distinct notes into the same time step, while one that is too fine increases sequence length and sparsity unnecessarily. To guide this decision, we analyzed the minimum inter-onset intervals ($\Delta t$) across the training corpus, obtaining the cumulative distributions functions shown in Figure ~\ref{fig:cdf}. Those distributions indicates that for each class more than 95\% of the consecutive notes are separated in time by more than 40ms.


In the following experiments, once a time resolution is chosen, charts with a minimum $\Delta t$ smaller than the chosen resolution are excluded from training. For instance, choosing a time resolution of 40 ms would result in the removal of roughly 1\% of the samples of \textit{Expert} difficulty.


\subsection{Baseline}
\label{sec:baseline}

Since no established benchmarks exist on \textit{Guitar Hero style} chart generation, a well defined baseline is essential to contextualize the results of more advanced models. The goal of this baseline is not to model the full charting process but to provide a clear reference for the \textit{note prediction capability} of a simple autoregressive model in the absence of temporal or audio information.

While the main proposed approach relies on a time-discretized formulation which introduces pad tokens to represent silent instants, this baseline intentionally removes any notion of time. The model is trained to predict notes in the order they appear in the chart, without representing their temporal placement either explicitly (as continuous offsets) or implicitly (through a fixed temporal grid). This simplifies the task considerably and can be interpreted as an approximate upper bound for unconditional note prediction accuracy.

The baseline is implemented as a decoder-only Transformer with approximately 6M parameters. Larger models were not explored, as initial experiments showed signs of overfitting. Training is performed on non-overlapping sequences extracted from each chart, focusing exclusively on the \textit{Expert} difficulty level.

\begin{figure}[h]
\centering
\includegraphics[width=1.0\linewidth]{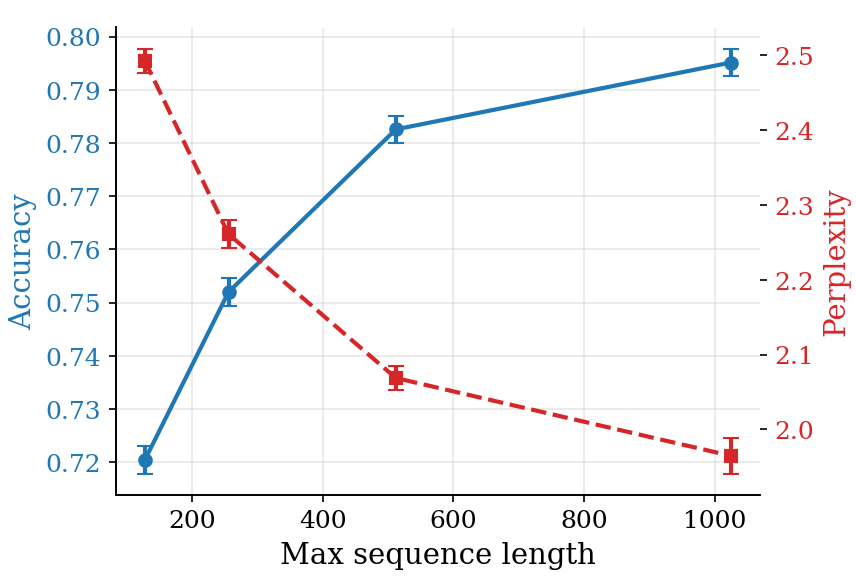}
\caption{Baseline performance in terms of perplexity and prediction accuracy as a function of the maximum context length. The x-axis corresponds to 128, 256, 512 and 1024 tokens, which roughly correspond to 15\,s, 30\,s, 60\,s, and the full song for the \textit{Expert} difficulty level.}
\label{perf_vs_seq}
\end{figure}

Figure~\ref{perf_vs_seq} shows validation performance in terms of perplexity and accuracy for different context lengths. Longer contexts consistently lead to better results, likely because musical structures and their repeating patterns become easier to capture with larger receptive fields.

\subsection{Audio-conditioned architecture}
\label{sec:architecture}

\begin{figure*}[htbp]
    \centering
    \includegraphics[width=0.7\textwidth]{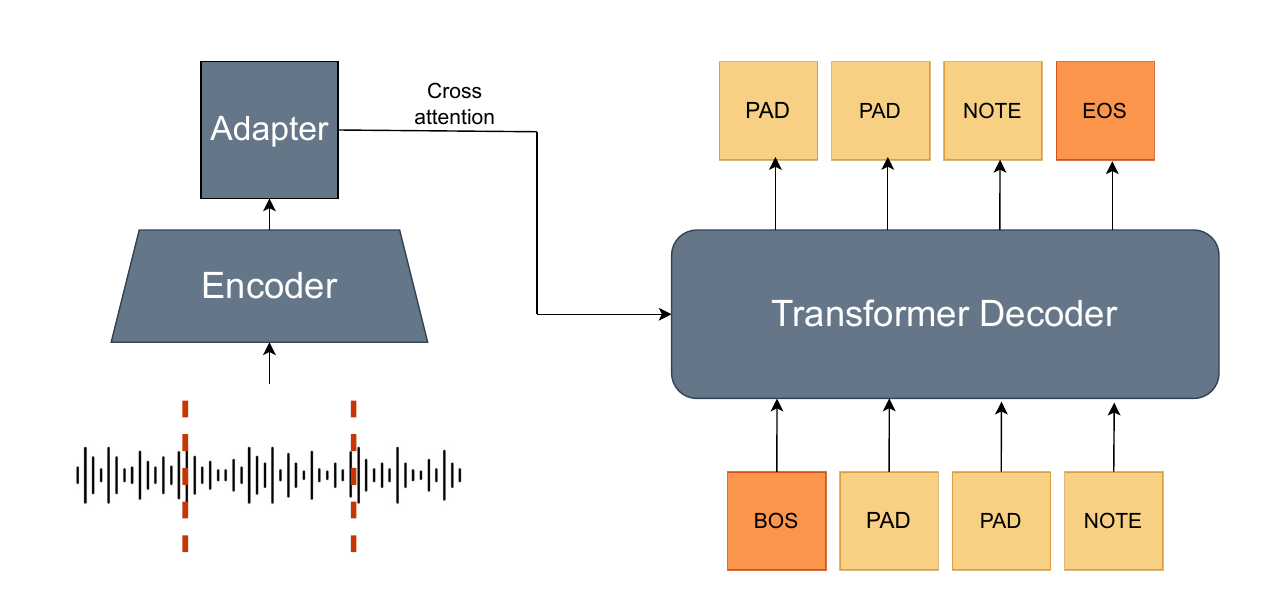}
    \caption{Overview of the proposed multimodal architecture. Audio is encoded in 30\,ms frames and processed through a pretrained Encodec encoder followed by a lightweight adapter. The resulting continuous representations are fused with the symbolic token sequence via cross-attention in a Transformer decoder.}
    \label{fig:architecture}
\end{figure*}

The audio-conditioned models operate on discretized time steps and follow an encoder-decoder design. The encoder processes the raw waveform 
$\mathbf{x} \in \mathbb{R}^{T_a \times F_s}$, 
where $T_a$ denotes the audio duration and $F_s$ the sampling rate.
A pretrained Encodec model \cite{défossez2022highfidelityneuralaudio} converts the waveform into a sequence of quantized audio codes
\[
\mathbf{z} \in \mathbb{N}^{T_e \times N_q},
\]
where $T_e = T_a \cdot F_{\text{enc}}$ is the number of encoder time steps and $N_q$ is the number of codebooks.
Each row $\mathbf{z}_t$ contains the discrete code indices corresponding to the audio segment at time $t$.

Each index is mapped to a learnable embedding vector, and the embeddings from all $N_q$ codebooks are summed to form a single conditioning vector per time step. 
This results in the final audio conditioning tensor
\[
\mathbf{c} \in \mathbb{R}^{T_e \times D},
\]
where $D$ is the embedding dimension.
A 1D convolutional adapter downsamples the audio representation to approximately match the temporal resolution of the token sequence, improving alignment during cross-attention.

The decoder is a Transformer-based autoregressive model that receives as input a sequence of previously generated note tokens 
$\mathbf{y}_{<t} \in \mathbb{R}^{T_y \times D_y}$,
where $T_y$ is the sequence length and $D_y$ the token embedding dimension.
Each decoder block employs pre-layer RMS normalization, SwiGLU activation, and multi-head cross-attention to the encoded audio features.
The model is trained to predict the next token $y_t$ conditioned on the history $\mathbf{y}_{<t}$ and the audio context $\mathbf{c}$.
Special boundary tokens (\texttt{<BOS>}, \texttt{<EOS>}) mark the beginning and end of each sequence.

\begin{table*}[t!]
\centering
\resizebox{\textwidth}{!}{%
\begin{tabular}{lccccccc}
\toprule
\textbf{Model} & \textbf{Time Res. (ms)} & \textbf{\# Params (M)} & 
\textbf{Perplexity (full)} & \textbf{Perplexity (non-pad)} & 
\textbf{Accuracy (full)} & \textbf{Accuracy (non-pad)} \\
\midrule
\textit{Baseline (no padding)} & -- & 6 & -- & $2.261 \pm 0.019$ & -- & $0.752 \pm 0.003$ \\
\midrule
Unconditional & 20 & 25 & \cellcolor{bestgreen}$1.621 \pm 0.006$ & $2.644 \pm 0.021$ & \cellcolor{bestgreen}$0.950 \pm 0.002$ & $0.715 \pm 0.002$ \\
Audio-conditioned & 20 & 25 & $1.688 \pm 0.009$ & \cellcolor{bestgreen}$2.342 \pm 0.022$ & $0.935 \pm 0.002$ & \cellcolor{bestgreen}$0.741 \pm 0.003$ \\
\midrule
Unconditional & 40 & 25 & \cellcolor{bestgreen}$1.579 \pm 0.018$ & $2.547 \pm 0.062$ & \cellcolor{bestgreen}$0.952 \pm 0.002$ & $0.723 \pm 0.007$ \\
Audio-conditioned & 40 & 25 & $1.703 \pm 0.005$ & \cellcolor{bestgreen}$2.226 \pm 0.024$ & $0.905 \pm 0.004$ & \cellcolor{bestgreen}$0.753 \pm 0.003$ \\
\bottomrule
\end{tabular}}
\caption{Comparison between unconditional and audio-conditioned models (25M parameters) at two time resolutions, including the non-discretized baseline with 256 tokens context length. Metrics are reported on the full sequence and on non-pad tokens only. Audio conditioning consistently improves non-pad performance. In green best metric for each time resolution group.}
\label{tab:audio_vs_uncond}
\end{table*}

\begin{table*}[t!]
\centering
\resizebox{\textwidth}{!}{%
\begin{tabular}{lccccccc}
\toprule
\textbf{Model} & \textbf{Time Res. (ms)} & \textbf{\# Params (M)} & 
\textbf{Perplexity (full)} & \textbf{Perplexity (non-pad)} & 
\textbf{Accuracy (full)} & \textbf{Accuracy (non-pad)} \\
\midrule
Audio-conditioned & 20 & 25 & $1.688 \pm 0.009$ & $2.342 \pm 0.022$ & $0.935 \pm 0.002$ & $0.741 \pm 0.003$ \\
Audio-conditioned & 20 & 227 & \cellcolor{bestgreen}$1.620 \pm 0.001$ & \cellcolor{bestgreen}$2.212 \pm 0.006$ & \cellcolor{bestgreen}$0.941 \pm 0.001$ & \cellcolor{bestgreen}$0.758 \pm 0.002$ \\
\midrule
Audio-conditioned & 40 & 25 & $1.703 \pm 0.005$ & $2.226 \pm 0.024$ & $0.905 \pm 0.004$ & $0.753 \pm 0.003$ \\
Audio-conditioned & 40 & 227 & \cellcolor{bestgreen}$1.670 \pm 0.007$ & \cellcolor{bestgreen}$2.169 \pm 0.034$ & \cellcolor{bestgreen}$0.913 \pm 0.003$ & \cellcolor{bestgreen}$0.761 \pm 0.005$ \\
\bottomrule
\end{tabular}}
\caption{Comparison between small (25M) and medium (227M) audio-conditioned models. Larger models achieve consistent improvements across both full-sequence and non-pad metrics. In green best metric for each time resolution group.}
\label{tab:scale}
\end{table*}

\subsection{Results}
\label{sec:results}

Model performance is evaluated using \textit{accuracy} and \textit{perplexity}, computed both on the \textit{full sequence} (including \textit{pad} tokens, representing silent time steps) and on the \textit{non-pad subset} corresponding to active note positions. 
This distinction is essential: high accuracy on the full sequence can result from trivially predicting pads, which account for about 90\% and 80\% of the tokens at 20\,ms and 40\,ms resolutions, respectively. 
The non-pad metrics therefore provide a more meaningful measure of the model's ability to capture musical content.

The results in Table~\ref{tab:audio_vs_uncond} show that the baseline model of Section~\ref{sec:baseline}, trained on direct note sequences without temporal discretization, achieves strong non-pad performance due to the simpler prediction setup. 
Among the baseline configurations, the version reported in the table corresponds to the model trained with a context window of 256 tokens, which roughly matches the average number of note events contained in a 30\,s chunk of music. 
This choice ensures a fair comparison with the discrete-time audio-conditioned models, for which each training sequence corresponds to a 30\,s audio segment. 
When time discretization and pad tokens are introduced, accuracy on note tokens naturally decreases as the task becomes more challenging and sequence length increases. 
The unconditional discrete-time models obtain higher scores on full-sequence metrics, largely because they can trivially predict the \textit{pad token} over most time steps, inflating overall accuracy and reducing perplexity without improving note prediction quality. 
In contrast, the audio-conditioned models consistently outperform unconditional ones on non-pad metrics for both 20\,ms and 40\,ms configurations, confirming that acoustic features provide valuable temporal cues. 
Notably, the audio-conditioned model at 40\,ms surpasses the non-discretized baseline in both non-pad accuracy and perplexity, demonstrating that integrating audio information not only compensates for the increased difficulty of the discretized setting but ultimately leads to superior predictive performance. 
This result is particularly important, as only the discretized audio-conditioned model possesses an explicit notion of time and can therefore be used to generate complete and temporally aligned charts from raw audio, unlike the baseline model which operates purely on symbolic note sequences.

As shown in Table~\ref{tab:scale}, scaling the audio-conditioned model from 25M to 227M parameters yields further gains in both full and non-pad metrics, indicating that additional capacity improves multimodal fusion and token-level prediction. 
While the 20\,ms configuration achieves higher accuracy, it doubles sequence length and inference cost without visible improvement in perceptual chart quality. 
The 40\,ms models thus represent the best trade-off between computational efficiency and output fidelity, being preferred for both training and generation.

All models are trained using the AdamW optimizer (weight decay 0.01) with a peak learning rate of $10^{-3}$ decayed to $10^{-4}$ by epoch 10. 
Each training sequence corresponds to a 30\,s audio segment. 
Pad-token losses are down-weighted by factors of 0.1 and 0.2 for the 20\,ms and 40\,ms settings, respectively. 
A dropout rate of 0.2 and audio augmentations are applied throughout. 
The audio encoder remains frozen, and all experiments are run on a single AMD MI300X GPU.

\section{Conclusion}
This work presented an end-to-end approach for transcribing audio into playable \textit{Guitar Hero} charts using a Transformer-based autoregressive model conditioned on audio representations. By adopting a time-discretized formulation, we demonstrated that it is possible to learn meaningful temporal and structural patterns directly from audio, even under strong sequence sparsity caused by pad tokens. The audio-conditioned model consistently outperformed the unconditional one, highlighting the importance of leveraging audio context for accurate note prediction and placement.

While this study focused on a single strategy, several promising extensions remain open such as two-stage training pipelines. Furthermore, this work focused solely on predicting note onsets and timing, without incorporating additional gameplay mechanics such as HOPOs, tap notes, or star power, which are essential for high quality charts. Future work could also explore conditioning on difficulty levels to enable controllable chart generation across skill tiers, as well as integrating more sophisticated tokenization strategies beyond fixed time discretization.

Finally, although the current training setup uses 30\,s audio chunks to balance context and compute, extending training to longer time windows would likely improve temporal consistency and chart coherence, at the cost of increased computational demand. Overall, this work establishes a strong baseline for automatic \textit{Guitar Hero} chart generation and provides an open-source foundation for further research.

\section{Acknowledgement}
Thanks to AMD and AMD Developer Cloud for the MI300X that made this work possible.

\textit{Guitar Hero} is a registered trademark of Activision Publishing, Inc.
This work is not affiliated with or endorsed by Activision.

\bibliographystyle{plain}
\bibliography{references}

@misc{lin2019generationmanialearningsemanticallychoreograph,
      title={GenerationMania: Learning to Semantically Choreograph}, 
      author={Zhiyu Lin and Kyle Xiao and Mark Riedl},
      year={2019},
      eprint={1806.11170},
      archivePrefix={arXiv},
      primaryClass={cs.SD},
      url={https://arxiv.org/abs/1806.11170}, 
}

@misc{amodei2015deepspeech2endtoend,
      title={Deep Speech 2: End-to-End Speech Recognition in English and Mandarin}, 
      author={Dario Amodei and Rishita Anubhai and Eric Battenberg and Carl Case and Jared Casper and Bryan Catanzaro and Jingdong Chen and Mike Chrzanowski and Adam Coates and Greg Diamos and Erich Elsen and Jesse Engel and Linxi Fan and Christopher Fougner and Tony Han and Awni Hannun and Billy Jun and Patrick LeGresley and Libby Lin and Sharan Narang and Andrew Ng and Sherjil Ozair and Ryan Prenger and Jonathan Raiman and Sanjeev Satheesh and David Seetapun and Shubho Sengupta and Yi Wang and Zhiqian Wang and Chong Wang and Bo Xiao and Dani Yogatama and Jun Zhan and Zhenyao Zhu},
      year={2015},
      eprint={1512.02595},
      archivePrefix={arXiv},
      primaryClass={cs.CL},
      url={https://arxiv.org/abs/1512.02595}, 
}

@misc{donahue2017dancedanceconvolution,
      title={Dance Dance Convolution}, 
      author={Chris Donahue and Zachary C. Lipton and Julian McAuley},
      year={2017},
      eprint={1703.06891},
      archivePrefix={arXiv},
      primaryClass={cs.LG},
      url={https://arxiv.org/abs/1703.06891}, 
}

@misc{vaswani2023attentionneed,
      title={Attention Is All You Need}, 
      author={Ashish Vaswani and Noam Shazeer and Niki Parmar and Jakob Uszkoreit and Llion Jones and Aidan N. Gomez and Lukasz Kaiser and Illia Polosukhin},
      year={2023},
      eprint={1706.03762},
      archivePrefix={arXiv},
      primaryClass={cs.CL},
      url={https://arxiv.org/abs/1706.03762}, 
}

@misc{radford2022robustspeechrecognitionlargescale,
      title={Robust Speech Recognition via Large-Scale Weak Supervision}, 
      author={Alec Radford and Jong Wook Kim and Tao Xu and Greg Brockman and Christine McLeavey and Ilya Sutskever},
      year={2022},
      eprint={2212.04356},
      archivePrefix={arXiv},
      primaryClass={eess.AS},
      url={https://arxiv.org/abs/2212.04356}, 
}

@misc{yi2023beatalignedspectrogramtosequencegenerationrhythmgame,
      title={Beat-Aligned Spectrogram-to-Sequence Generation of Rhythm-Game Charts}, 
      author={Jayeon Yi and Sungho Lee and Kyogu Lee},
      year={2023},
      eprint={2311.13687},
      archivePrefix={arXiv},
      primaryClass={cs.LG},
      url={https://arxiv.org/abs/2311.13687}, 
}

@article{10.5555/177910.177914,
author = {Gage, Philip},
title = {A new algorithm for data compression},
year = {1994},
issue_date = {Feb. 1994},
publisher = {R \& D Publications, Inc.},
address = {USA},
volume = {12},
number = {2},
issn = {0898-9788},
journal = {C Users J.},
month = feb,
pages = {23–38},
numpages = {16}
}

@misc{défossez2022highfidelityneuralaudio,
      title={High Fidelity Neural Audio Compression}, 
      author={Alexandre Défossez and Jade Copet and Gabriel Synnaeve and Yossi Adi},
      year={2022},
      eprint={2210.13438},
      archivePrefix={arXiv},
      primaryClass={eess.AS},
      url={https://arxiv.org/abs/2210.13438}, 
}

\clearpage

\onecolumn
\appendix
\section{Dataset}
\label{appendix_dataset}

\begin{figure}[h]
\centering
\includegraphics[width=0.8\linewidth]{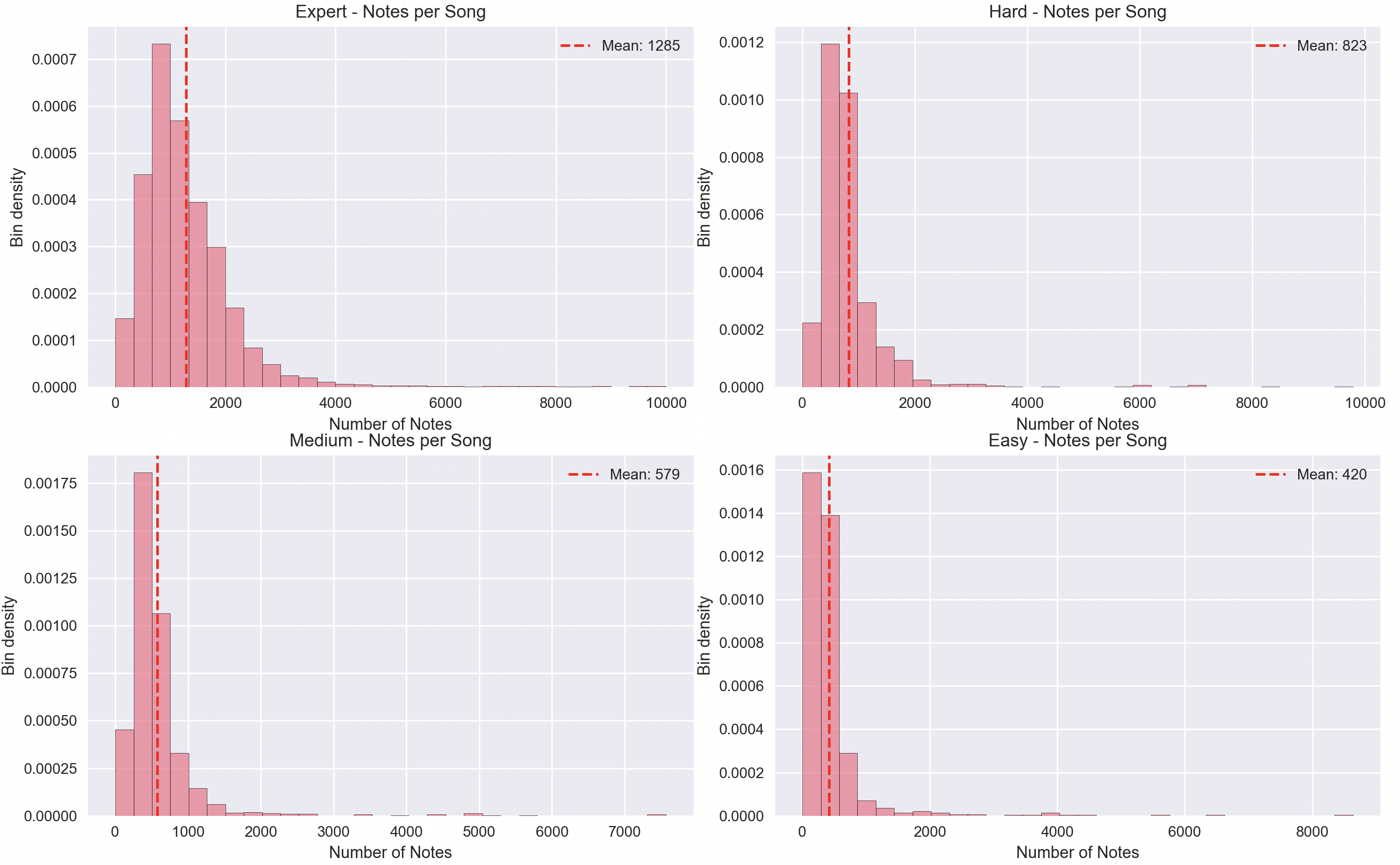}
\caption{Bin density of the number of notes per song for each difficulty level.}
\label{bin_density}
\end{figure}

\begin{flushleft}
The average song duration is the same for each difficulty class. Higher difficulty levels contain more notes per song as seen in Figure \ref{bin_density}.
\end{flushleft}

\begin{figure}[h]
\centering
\includegraphics[width=0.5\linewidth]{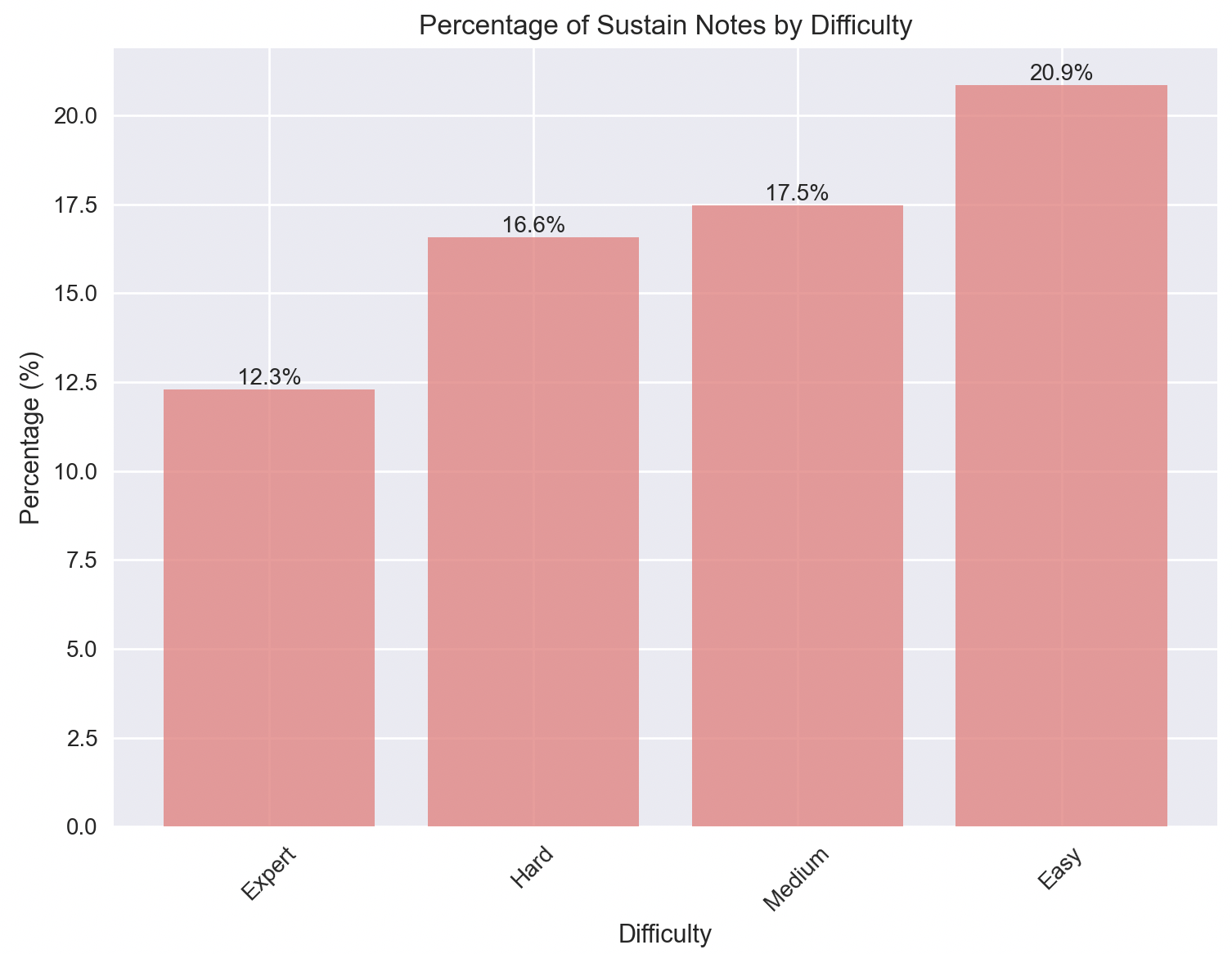}
\caption{Percentage of sustain notes per difficulty level.}
\label{sustain_notes}
\end{figure}

\begin{flushleft}
Interestingly, in Figure \ref{sustain_notes} can be seen that lower difficulty charts have an higher proportion of sustain notes, whereas higher difficulty levels favor a denser arrangement of tap notes, reflecting a design choice to increase challenge through note density rather than sustain duration.
\end{flushleft}

\begin{figure}[h]
\centering
\includegraphics[width=0.8\linewidth]{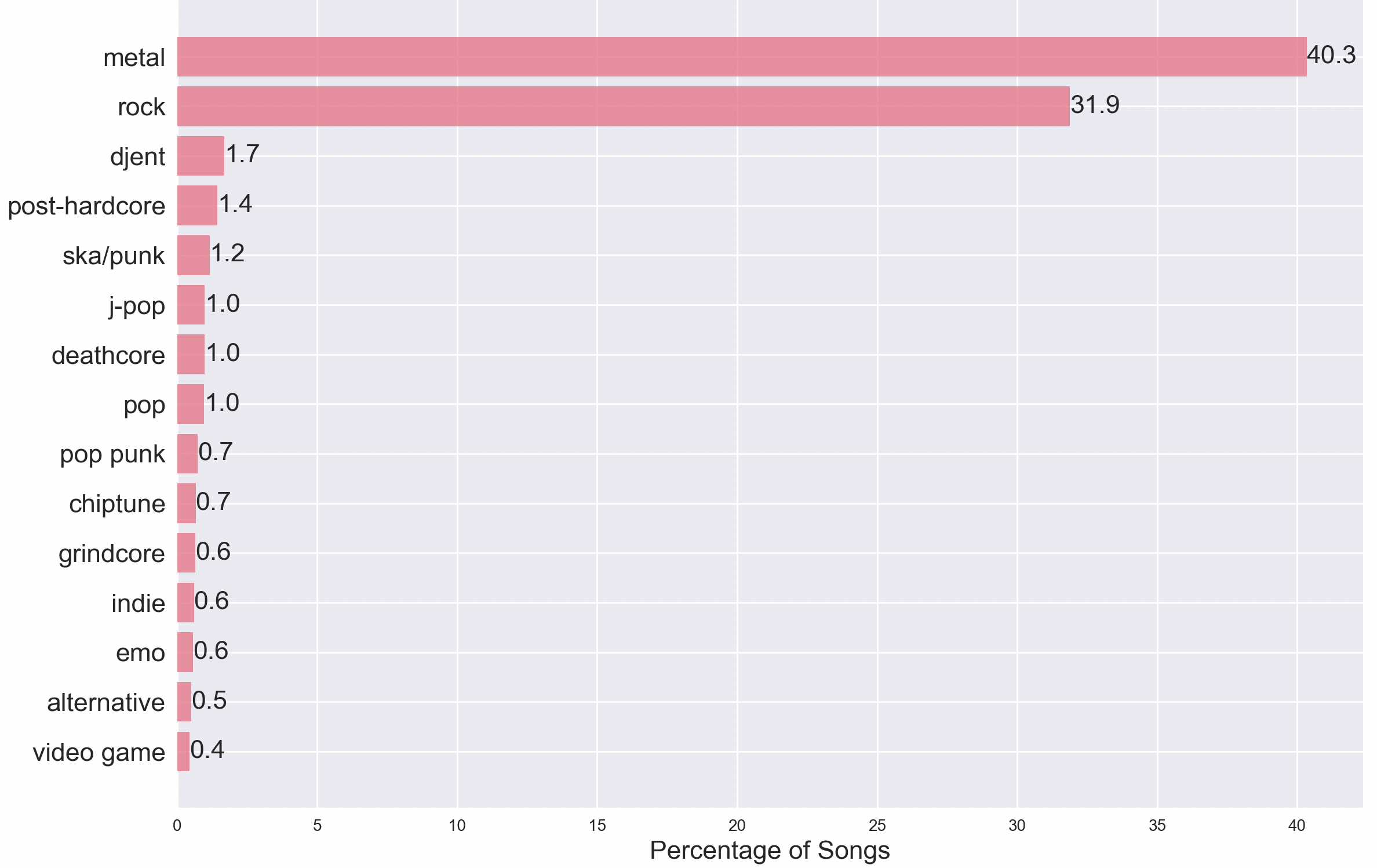}
\caption{Percentage of top-15 most frequent song genres.}
\label{genres}
\end{figure}

\begin{flushleft}
Finally, the dataset is mostly made up of metal and rock subgenres as we can see in Figure \ref{genres} where subgenres containing the word \emph{rock} or \emph{metal} have been grouped together.
\end{flushleft}

\section{Tokenization}
\label{appendix_tokenization}

\begin{figure}[h]
\centering
\includegraphics[width=0.8\linewidth]{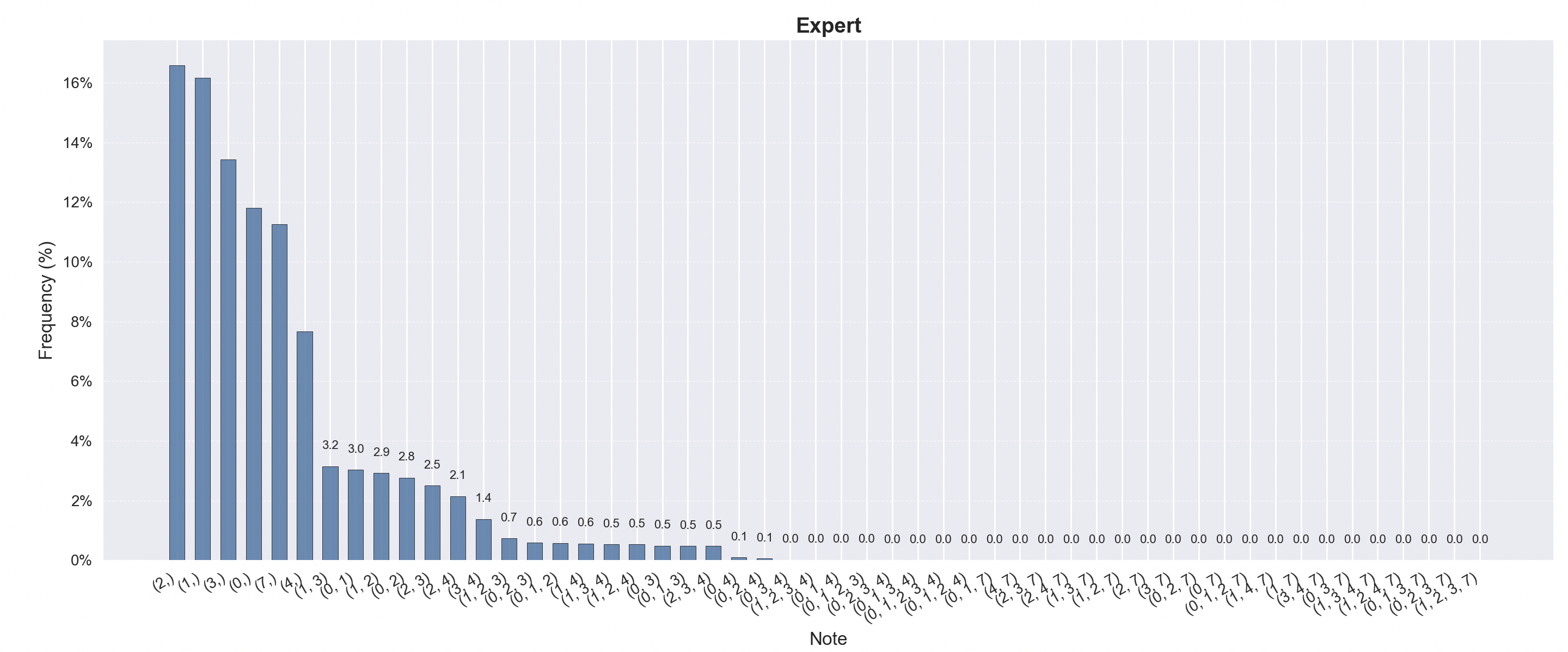}
\caption{Notes distribution for the Expert class. (The right tail of notes is present less that 0.1\%)}
\label{notes_expert}
\end{figure}

\begin{figure}[h]
\centering
\includegraphics[width=0.8\linewidth]{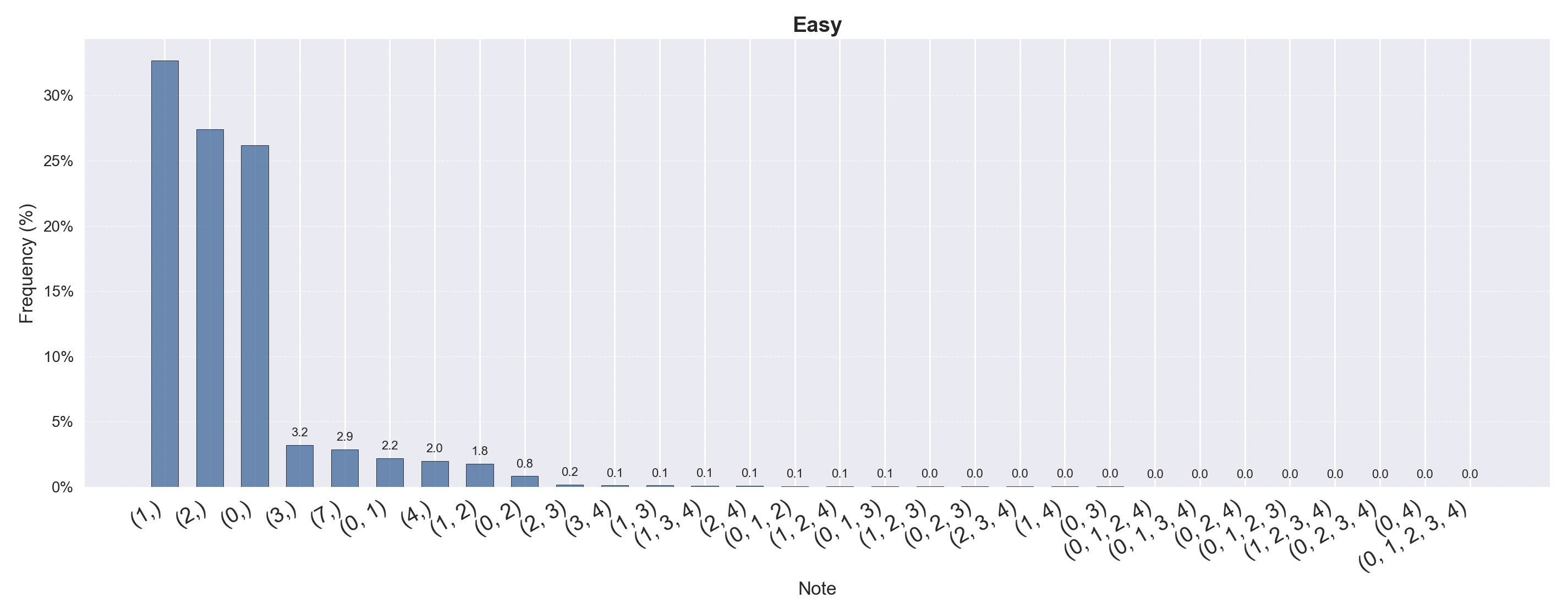}
\caption{Notes distribution for the Easy class. (The right tail of notes is present less that 0.1\%)}
\label{notes_easy}
\end{figure}

\begin{flushleft}
Each Guitar Hero note or chord is represented as a discrete token that uniquely encodes the combination of fret buttons being pressed. The guitar controller consists of five colored fret buttons (green, red, yellow, blue, and orange) which are indexed from $0$ to $4$, respectively. An additional index, $7$, represents an \emph{open note}. The tokenizer enumerates all possible non-empty combinations of these six elements (five frets plus the open note), generating 63 distinct button configurations in total. Each configuration is assigned a unique integer identifier, referred to as a \emph{token ID}.

The Expert class contains the largest amount of unique token IDs. Each class has a tail of low probability token (< 0.1\%), as seen in Figures \ref{notes_easy}, \ref{notes_expert} and similarly for Hard and Medium.
\end{flushleft}

\section{Multimodal Training}
\label{appendix_multimodal}

\begin{figure}[h]
\centering
\includegraphics[width=0.8\linewidth]{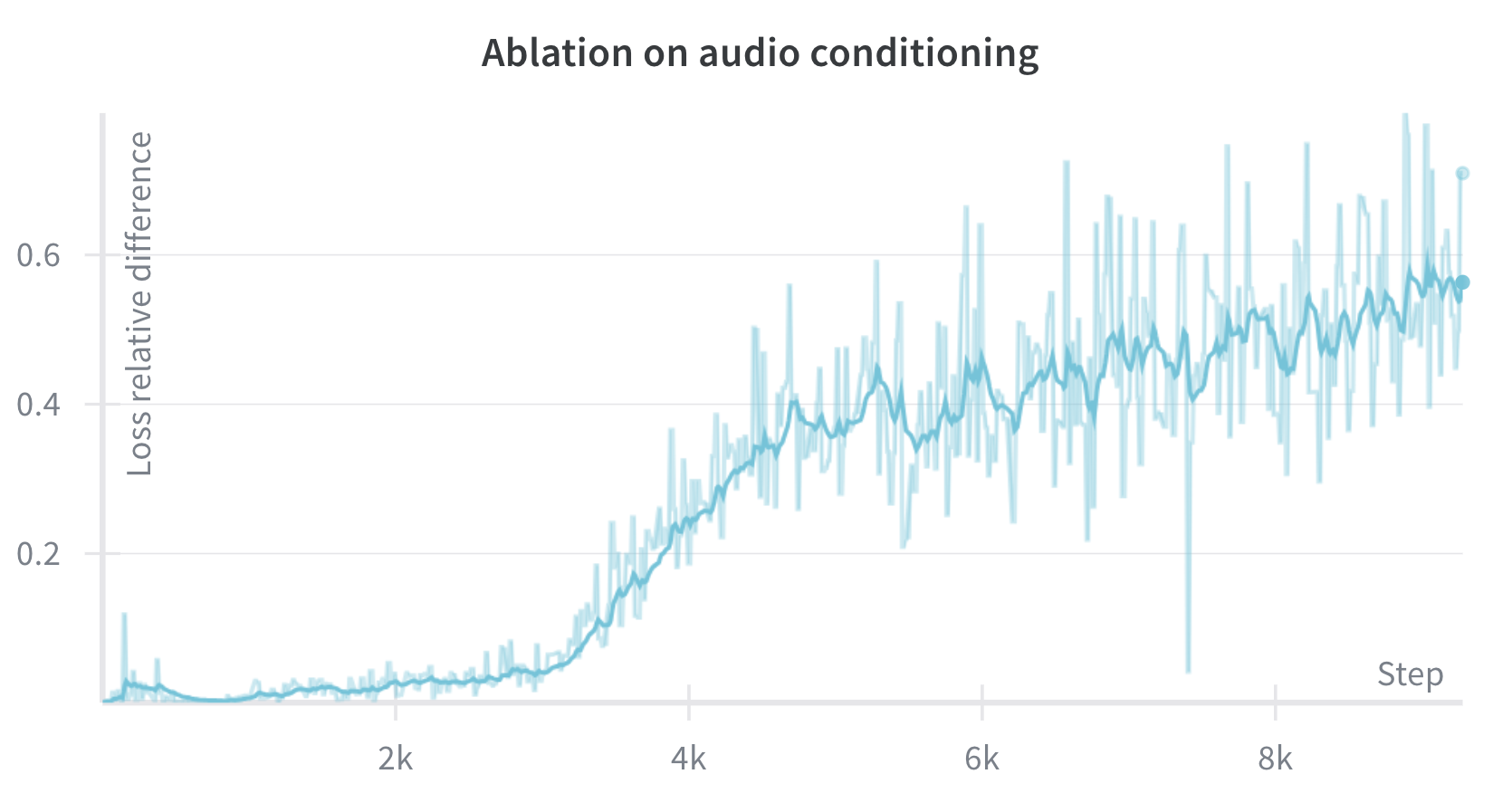}
\caption{Relative difference in loss between the standard forward pass and a forward pass with permuted audio samples during training. A growing gap indicates that the model progressively learns to exploit the audio modality rather than relying solely on the symbolic input.}
\label{ablation}
\end{figure}

The presence of a cross-attention module in the Transformer decoder does not necessarily guarantee that the model will effectively exploit the audio conditioning. To assess the actual contribution of the audio modality during training, we measure the relative difference between the standard forward pass and a forward pass in which the audio samples within the batch are randomly permuted. This ablation reveals whether the model relies on the audio or primarily on the autoregressive note context. As shown in Figure \ref{ablation} for the 25M-parameters model with a 20\,ms time resolution, the network initially learns to model the task using only the input note tokens, and only after approximately 2k training steps it begins to effectively leverage the audio information.

\end{document}